\documentclass{PoS}
\def\npb#1#2#3{Nucl.\ Phys.\ {\bf B#1} (#2) #3}
\def\plb#1#2#3{Phys.\ Lett.\ {\bf B#1} (#2) #3}
\def\eq#1{eq.~(\ref{#1})}
\newcommand{\bea}{\begin{eqnarray}}
\newcommand{\eea}{\end{eqnarray}}
\newcommand{\beq}{\begin{eqnarray}}
\newcommand{\eeq}{\end{eqnarray}}
\newcommand{\pslash}{p \!\!\!\!/}
\title{Vacuum expectation value of $A^2$ from LQCD.}

\ShortTitle{$\langle A^2\rangle$ from LQCD in Landau gauge}

\author{\speaker{O. P\`ene}\thanks{O.P. wishes to thank the organisers
for the very pleasant meeting in such a wonderful city.}\\
	Laboratoire de Physique Th\'eorique\footnote{Unit\'e Mixte de Recherche 8627 du Centre National de
la Recherche Scientifique},\\
CNRS et Universit\'e  Paris-Sud XI, B\^atiment 210, 91405 Orsay Cedex,
France\\
	E-mail: \email{olivier.pene@th.u-psud.fr}}

\author{Benoit Blossier, Ph. Boucaud, A. Le Yaouanc, J.P. Leroy, J. Micheli\\
	Laboratoire de Physique Th\'eorique\\
CNRS et Universit\'e  Paris-Sud XI, B\^atiment 210, 91405 Orsay Cedex,
France}	
	\author{ M.Brinet\\
  Laboratoire de Physique Subatomique et de Cosmologie, CNRS/IN2P3/UJF, \\
53, avenue des Martyrs, 38026 Grenoble, France}

\author{M. Gravina\\
Department of Physics,  University of Cyprus, P.O. Box 20537, 1678 Nicosia,\\
Cyprus and Computation-based Science and Technology Research Center}

\author{F. De Soto\\
	Dpto. Sistemas F\'isicos, Qu\'imicos y Naturales, \\
Universidad Pablo de Olavide, 41013 Sevilla, Spain}

\author{ Z. Liu\\
	DAMTP, University of Cambridge, \\
Wilberforce Road, Cambridge CB3 0WA, United Kingdom}
	
	\author{ V. Morenas\\
Laboratoire de Physique Corpusculaire, Universit\'e Blaise Pascal, CNRS/IN2P3 \\
63000 Aubi\`ere Cedex, France}
	
	\author{ K. Petrov\\
	SPhN-IRFU, CEA Saclay, F91191 Gif sur Yvette, France}
	
		\author{ J.~Rodr\'iguez-Quintero\\
	Dpto. F\'isica Aplicada, Fac. Ciencias Experimentales,\\
Universidad de Huelva, 21071 Huelva, Spain}
	
\abstract{We argue from LQCD that there is a non vanishing v.e.v 
of $A_a^\mu A^a_\mu$ in QCD in the Landau gauge. We use operator product
expansion to provide a clear definition of $A_a^\mu A^a_\mu$ and extract a number
both in the quenched and unquenched case. }

\FullConference{The many faces of QCD\\
		November 2-5, 2010\\
		Gent Belgium}

\begin{document}

%============================================================================
\section{Introduction}
``Natura abhorret vaccum" was an antique saying. Modern science has strangely
supported this intuition when discovering that vacuum is very far from being
empty. There are quantum fluctuations, virtual  $e^+e^-$ and $q \bar q$ pairs
and vacuum expectation values (v.e.v). Our target here is the v.e.v of 
$A^2\equiv A_a^\mu A^a_\mu$  in Landau gauge. 

Is it legitimate to speak about a gauge dependent property of a gauge invariant
vacuum ? Yes, of course, once we have understood that there are fields in the
vacuum. These fields may be scrutinized in different gauges, the appearance will
differ, the physics we learn will vary, and nevertheless the vacuum is gauge
invariant. It is absolutely similar to the well known fact that a physical
process, say some reaction, is decomposed differently into Feynman diagrams
according to the chosen gauge, its appearance will depend on the gauge although
the process is gauge invariant and the cross section will not depend on the
gauge.  

$A^2$ is certainly not a gauge invariant quantity, but in Landau gauge it is 
invariant for infinitesimal gauge tranformations, and consequently, for BRST
transformations. 

Why do we care about $\langle A^2 \rangle$ ? It is useful 
\begin{itemize}
\item in order to understand the infrared properties of QCD. It is enough
to quote an incomplete list of studies of QCD in the infrared which use 
$\langle A^2 \rangle$~\cite{Gubarev:2000ym}-\cite{Blossier:2010ph}, many of the
authors being in the audience;
\item in order to identify correctly non-perturbative corrections to
renormalisation constants;
\item in order to test and calibrate on a rather extensively computable case Operator
Product Expansion : we find larger non-perturbative corrections than 
usually expected. 
\end{itemize}

%============================================
\section{How to define $\langle A^2 \rangle$ ?}
A naive estimate of $\langle A^2 \rangle$ produces an ultraviolet divergent
quantity $\propto a^{-2}$ ($a$ being the lattice spacing). Indeed we are interested
in infrared modes and we must define a scale $\mu$ which, grossly speaking,
separates the high and low modes.
A way to perform rigorously the distinction between UV and IR modes is
to use {\bf Operator Product Expansion (OPE)}. One must then define a
renormalisation scheme and a renormalisation scale.

%============================================
\section{Wilson Operator expansion}

A momentum dependent quantity $Q(p^2)$ with vacuum quantum numbers 
can be expanded in inverse powers of $\mu^2$
\bea\label{eq:OPE}
Q(p^2)=   Q_{\rm pert}(p^2,\mu^2)  + C^Q_{\rm wilson}(p^2,\mu^2) 
\langle A^2(\mu^2)\rangle  + ....
\eea
where $Q_{\rm pert}(p^2,\mu^2)$ and $C^Q_{\rm wilson}(p^2,\mu^2)\sim 1/\mu^2$ are 
series in $\alpha(\mu^2)$ computable in perturbative QCD. $Q$ can be the strong
coupling constant, the quark field renormalisation constant, other
renormalisation constants, etc. The coefficients $C^Q_{\rm wilson}$ are
often called Wilson coefficients.

OPE has been extensively used in phenomenology since the pioneering work by 
SVZ~\cite{SVZ}\footnote{And we have the pleasure to have ``Z'' in the audience.}.

%%%%%%%%%%%%%%%%%%%%%%%%%%%%%%%%%%%%%%%%%%%%%%%%%%%%%%%%%%%%%%%
\subsection{Criteria to check that we really measure 
$\langle A^2(\mu^2)\rangle$.}
\begin{itemize}
\item
The use of OPE is criticized under the argument that it is difficult to 
distinguish higher orders in the perturbative series (which are logarithmically
suppressed $\propto 1/\log(p^2)$) from OPE contributions which are power
suppressed ($\propto 1/p^2$) and that perturbative renormalons mimic a
condensate, being
also $\propto 1/p^2$.  This issue has been carefully studied 
by Martinelli and Sachrajda (MS) in~\cite{Martinelli:1996pk}. They show, using a
specific model, that
when subtracting with proper factors two quantities, say in our case $Z_q$ and
$\alpha_T$, renormalon ambiguities cancel. One must still check that the
renormalon-free
neglected terms in the perturbative expansion are not larger than the
condensate.    
 The criterium proposed by MS is that we must check that the last perturbative
 contribution considered  (say the order $\alpha^3$ or $\alpha^4$ depending on
 what is available) is significantly smaller than the power correction.
 This criterium has been checked in~\cite{Blossier:2010vt} concluding that 
 the MS rule was satisfied on the full momentum range which is used. 

\item $\langle A^2(\mu^2)\rangle $ is not a lattice artefact. It is
 a quantity defined in the continuum limit. 
Therefore we must check that what we interpret as  
$\langle A^2(\mu^2)\rangle$ does not depend significanlty on the lattice spacing.

\item We must also check that $\langle A^2(\mu^2)\rangle$ derived from different
 quantities are consistent provided we use the same renormalisation scheme
  and scale.
  
\item   $\langle A^2(\mu^2)\rangle$ depends on the vacuum and there is no
reason for the values of $\langle A^2(\mu^2)\rangle$ extracted from $N_f=0$ and 
$N_f=2$ lattice data to agree.  
\end{itemize}

%%%%%%%%%%%%%%%%%%%%%%%%%%%%%%%%%%%%%%%%%%%%%%%%%%%%%%%%%%%%%%%
\subsection{About Wilson coefficients}
All Wilson coefficients $C^Q_{\rm wilson}(p^2,\mu^2)$  are of the type 
\bea\label{eq:CW}
C^Q_{\rm wilson}(p^2,\mu^2)= d^Q{\rm tree}\;
g^2(\mu^2)\;\frac{1+\cal{O}(\alpha)}{p^2} \quad {\rm where}\nonumber \\ 
d^Q{\rm tree}= \frac 1 {12}\;\; {\rm for}\;\;\; Q=Z_q,\qquad  d^Q{\rm tree} = \frac 9
{32} \;\; {\rm for}\;\;\;Q=\alpha_T.
\eea
From \eq{eq:OPE} and \eq{eq:CW} we see that there is always the same factor
$\langle g^2(\mu^2) A^2(\mu^2)\rangle_{\rm \overline MS}$ and we will therefore
 give the fitted values of the condensate as  
 $\langle g^2(\mu^2) A^2(\mu^2)\rangle_{\rm \overline {MS}}$.
Indeed, in the following, we choose to renormalise the Wilson coefficient such that the
local operator $A^2(\mu^2)$ is in the $\rm \overline {MS}$ scheme, whichever 
prescription  we take for  the perturbative part of $Q$.

In practice one can show that the best and most general fitting formula is:
\bea\label{eq:fact}
Q(p^2)=  Q_{\rm pert}(p^2,\mu^2) \left (1 + 
\frac{C^Q_{\rm wilson}(p^2,\mu^2)}{Q_{\rm pert}(p^2,\mu^2)}\;\;
\langle A^2(\mu^2)\rangle_{\rm \overline MS} \right )
\eea
At leading logarithm for the non-perturbative correction, 
\bea \label{eq:LL}
&&\frac {C^Q_{\rm wilson}(p^2,\mu^2)}{Q_{\rm pert}(p^2,\mu^2)}\;\; 
\langle A^2(\mu^2)\rangle_{\rm \overline MS} = d^Q{\rm tree}\;\;
\langle g^2(\mu^2) A^2(\mu^2)\rangle_{\rm \overline MS} \left(\frac
{\alpha(p^2)}{\alpha(\mu^2)}
\right)^e\qquad\nonumber \\
&&{\rm where}\qquad e=\frac{9}{44- \frac{8 \, N_f}3 }.
\eea
Notice that $e$ is small. Therefore the corrective
factor in \eq{eq:LL} is almost scale invariant. The exponent is the same for all
quantities. In \eq{eq:LL} the only term which depends on the measured quantity is 
$d^Q{\rm tree}$. 

%%%%%%%%%%%%%%%%%%%%%%%%%%%%%%%%%%%%%%%%%%%%%%%%%%%%%%%%%%
\subsection{How to compute  $\langle A^2(\mu^2)\rangle_{\rm \overline
MS}$   from LQCD ?}

\begin{itemize}
\item Compute $Q_{\rm latt}$ as measured from LQCD.
\item Correct for hypercubic lattice artefacts i.e. artefacts related to the
 hypercubic geometry.
\item Fit according to 
$$
Q(p^2)=  Q_{\rm pert}(p^2,\mu^2) \left (1 + 
\frac{C^Q_{\rm wilson}(p^2,\mu^2)}{Q_{\rm pert}(p^2,\mu^2)}\;\;
\langle A^2(\mu^2)\rangle_{\rm \overline MS} \right ) + c_{a2p2}\; a^2p^2
$$ where $C_{a2p2}\, a^2p^2$ is a simple model for the non-hypercubic 
remaining lattice artefacts, which turns out to give a good result, as we shall
illustrate below. 
\end{itemize}
Let us now apply this strategy.

%==============================================
\section{The strong coupling constant}
There are many ways to define the strong coupling constant. We will 
use~\cite{von Smekal:1997is,Boucaud:2008gn} what we call the ``Taylor coupling
constant" which, thanks to Taylor's theorem~\cite{Taylor:1971ff}, is only 
dependent on the gluon and ghost propagators.

We will use configurations with Wilson twisted quarks (Nf=2) from the ETM 
collaboration and compare them to quenched configurations (Nf=0).  

\subsection{Some definitions}
The gluon propagator ($G^{(2)}$) and the ghost propagator ($F^{(2)}$) are defined
as follows, $G(p^2,\Lambda)$ and $F(p^2,\Lambda)$ being named the ``dressing
functions":
\bea
\left( G^{(2)} \right)_{\mu \nu}^{a b}(p^2,\Lambda) &=& \frac{G(p^2,\Lambda)}{p^2} \ \delta_{a b} 
\left( \delta_{\mu \nu}-\frac{p_\mu p_\nu}{p^2} \right) \ ,
%\langle \widetilde{A^a_\mu}(-p) \widetilde{A^b_\nu}(p) \rangle \ ,
\nonumber \\
\left(F^{(2)} \right)^{a,b}(p^2,\Lambda) &=& - \delta_{a b} \ \frac{F(p^2,\Lambda)}{p^2} \ ;
\eea
$\Lambda$ being some regularisation parameter: $\Lambda=a^{-1}(\beta)$ if, for 
instance, we specialise to lattice regularisation. The renormalised dressing
functions, $G_R$ and $F_R$ are defined through :
\begin{flushleft}%\begin{subequations}
\bea \label{bar}
G_R(p^2,\mu^2)\ &= \ \lim_{\Lambda \to \infty} Z_3^{-1}(\mu^2,\Lambda) \
G(p^2,\Lambda)\nonumber\\
F_R(p^2,\mu^2)\ &= \ \lim_{\Lambda \to \infty}
\widetilde{Z}_3^{-1}(\mu^2,\Lambda)\ F(p^2,\Lambda) \ ,
\eea%\end{subequations}
\end{flushleft}
\noindent with renormalisation condition
\bea\label{bar2}
G_R(\mu^2,\mu^2)=F_R(\mu^2,\mu^2)=1 \ .
\eea
Now, we will consider the ghost-gluon vertex which could be non-perturbatively
obtained through  a three-point Green function, defined by two ghost and one
gluon fields,  with amputated legs after dividing by two ghost and one gluon 
propagators. This vertex can be written quite generally as:

\beq\label{defGamma}
\widetilde{\Gamma}^{abc}_\nu(-q,k;q-k) = 
%\ghvertex =
i g_0 f^{abc} 
\left( q_\nu H_1(q,k) + (q-k)_\nu H_2(q,k) \right) \ ,
\eeq
The vertex renormalisation constant is defined as
\beq\label{MOMT}
\left.(H^R_1(q,k) +  H^R_2(q,k))\right\vert_{q^2=\mu^2} = 
\lim_{\Lambda \to \infty}\widetilde{Z}_1(\mu^2,\Lambda)\left.(H_1(q,k;\Lambda) 
+  H_2(q,k;\Lambda))\right\vert_{q^2=\mu^2} =1,
\eeq

The renormalised coupling constant is defined as
\beq\label{g2R}
g_R(\mu^2) &=& \lim_{\Lambda \to \infty} \ \widetilde{Z}_3(\mu^2,\Lambda) Z_3^{1/2}(\mu^2,\Lambda) g_0(\Lambda^2)   
\left. \left(  H_1(q,k;\Lambda) + H_2(q,k;\Lambda) 
\rule[0cm]{0cm}{0.5cm}  \right) \right|_{q^2 \equiv \mu^2} 
\nonumber \\
&=&  \ \lim_{\Lambda \to \infty} g_0(\Lambda^2) \ 
\frac{Z_3^{1/2}(\mu^2,\Lambda^2)\widetilde{Z}_3(\mu^2,\Lambda^2)}{ \widetilde{Z}_1(\mu^2,\Lambda^2)} \ .
\eeq
 Now we choose a special kinematics where the incoming ghost momentum vanishes.
Taylor's theorem states that $H_1(q,0;\Lambda) +  H_2(q,0;\Lambda)$ is equal to
1 in full QCD for  any value of $q$. Therefore, the renormalisation condition
\eq{MOMT} implies $\widetilde{Z}_1(\mu^2)=1 $ and then 
\beq\label{alpha} 
\alpha_T(\mu^2) \equiv \frac{g^2_T(\mu^2)}{4 \pi}=  \ \lim_{\Lambda \to \infty} 
\frac{g_0^2(\Lambda^2)}{4 \pi} G(\mu^2,\Lambda^2) F^{2}(\mu^2,\Lambda^2) \ ;
\eeq
which only depends on the propagators.

\subsection{Perturbative running of $\alpha_T$}
The four-loops  expression for the coupling constant in the Taylor
scheme as a function of $\Lambda_T$ ($\Lambda_{\rm QCD}$ in this scheme) 
is given by~\cite{Chetyrkin00,Chetyrkin:2004mf,Boucaud:2008gn}:
%\begin{align}
\beq
  \label{betainvert}
%  \begin{split}
      \alpha_T(\mu^2) &=& \frac{4 \pi}{\beta_{0}t}
      \left(1 - \frac{\beta_{1}}{\beta_{0}^{2}}\frac{\log(t)}{t}
     + \frac{\beta_{1}^{2}}{\beta_{0}^{4}}
       \frac{1}{t^{2}}\left(\left(\log(t)-\frac{1}{2}\right)^{2}
     + \frac{\widetilde{\beta}_{2}\beta_{0}}{\beta_{1}^{2}}-\frac{5}{4}\right)\right)  \\
     &+& \frac{1}{(\beta_{0}t)^{4}}
 \left(\frac{\widetilde{\beta}_{3}}{2\beta_{0}}+
   \frac{1}{2}\left(\frac{\beta_{1}}{\beta_{0}}\right)^{3}
   \left(-2\log^{3}(t)+5\log^{2}(t)+
\left(4-6\frac{\widetilde{\beta}_{2}\beta_{0}}{\beta_{1}^{2}}\right)\log(t)-1\right)\right)
\nonumber
%   \end{split}
%\end{align}
\eeq
where $t=\ln \frac{\mu^2}{\Lambda_T^2}$ and coefficients are
\beq\label{betacoefs}
\beta_0 &=&  11 - \frac 2 3 N_f \ , \ \
\beta_1 = \overline{\beta}_1 = 102 - \frac{38} 3 N_f
\nonumber \\
\widetilde{\beta}_2 &=& 3040.48 \ - \ 625.387 \ N_f \ + \ 19.3833 \ N_f^2
\nonumber \\
\widetilde{\beta}_3 &=& 100541 \ - \ 24423.3 \ N_f \ + \ 1625.4 \ N_f^2 \ - \ 27.493 \ N_f^3
\ ,
\eeq
$\Lambda_T$ is converted into $\Lambda_{\overline{\rm MS}}$ by
\beq\label{ratTMS}
\frac{\Lambda_{\overline{\rm MS}}}{\Lambda_T} \ = \ e^{\displaystyle -\frac{c_1}{2 \beta_0}} \ = \ 
e^{\displaystyle - \frac{507-40 N_f}{792 - 48 N_f}}
\ .
\eeq

\subsection{The non-perturbative contribution to $\alpha_T$}

It is easy to see that the dominant non-perturbative contribution to 
$\alpha_T(\mu^2)$ is the condensate $\langle A^2 \rangle$. 
To take it into account in our fits we will need the Wilson coefficient
$C^{\alpha_T}_{\rm wilson}$. This is possible up to three loops thanks 
to~\cite{Chetyrkin:2009kh} where the $\langle A^2 \rangle$ correction to the
ghost and gluon propagators have been computed. We will only write here 
the result at leading logarithm~\cite{Boucaud:2008gn}.
\beq\label{ap-alphahNP}
\alpha_T(\mu^2)
\ = \
\alpha^{\rm pert}_T(\mu^2)
\ 
\left( 
 1 + \frac{9}{\mu^2} 
\left( 
\frac{\ln\frac{\mu^2}{\Lambda^2_{QCD}}}{\ln\frac{\mu_0^2}{\Lambda^2_{QCD}}}
\right)^{-9/(44-8\,N_f/3)}
\frac{g^2_T(\mu_0^2) \langle A^2 \rangle_{R,\mu_0^2}} {4 (N_C^2-1)}
\right) \ ,
\eeq
where $\mu_0$ is our reference renormalisation scale which we take to be 
10 GeV.

\subsection{Results of the lattice simulation}
\begin{center}
\begin{tabular}{cc}
 \begin{minipage}{0.60\linewidth}
\hskip -2cm \includegraphics[width=11cm]{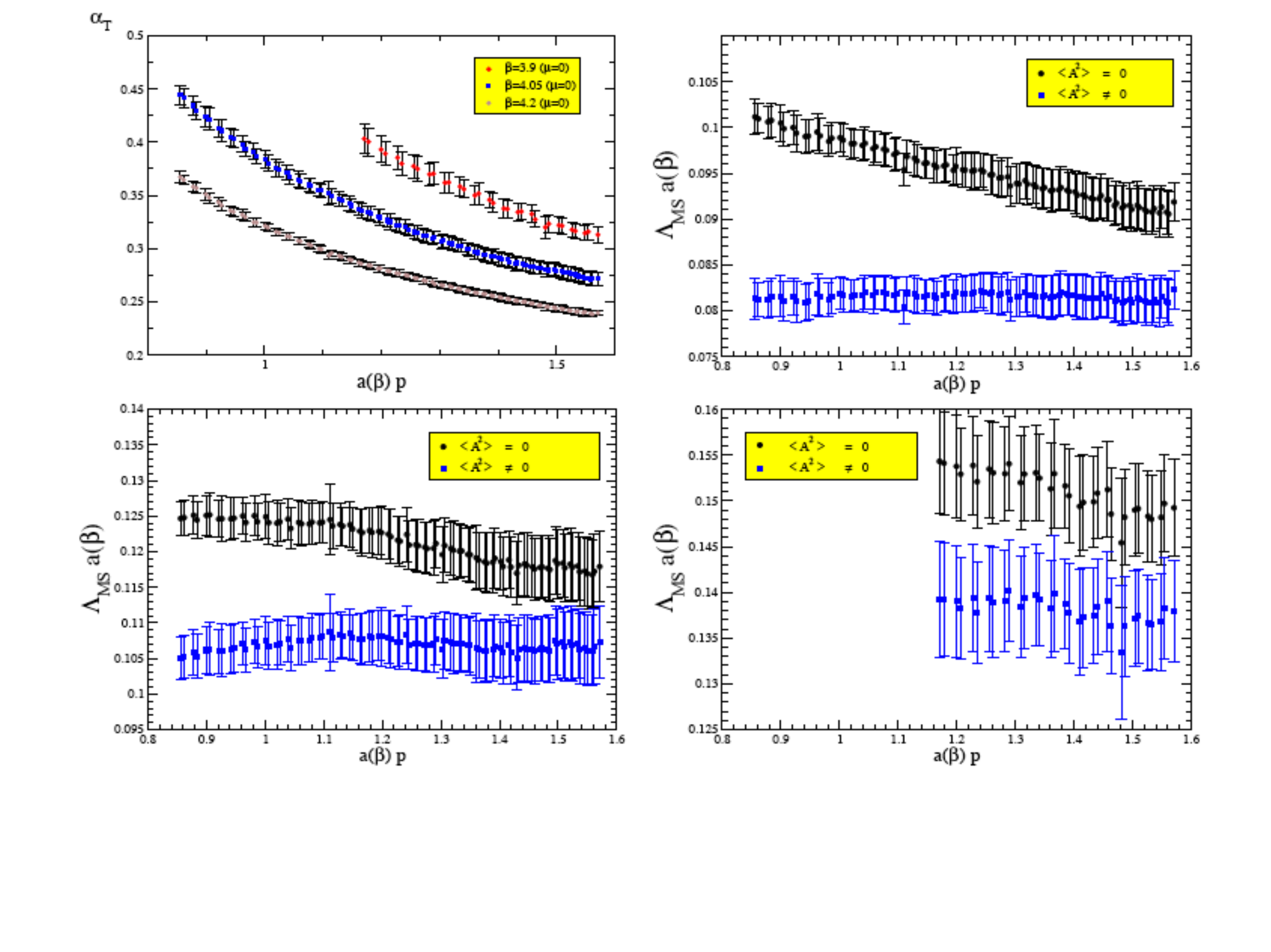}
 \end{minipage}
 \begin{minipage}{0.40\linewidth}
We perform the lattice calculation of $\alpha_T(\mu^2)$ from eq. (\ref{alpha}),
using $N_f=2$ with twisted dynamical quarks from the ETMC collaboration. We
eliminate the hypercubic lattice artefacts. The results depend on the dynamical
quark mass. For three values of the lattice spacing (labelled by $\beta$)
we extrapolate to zero mass leading to the three sets of data in the upper-left 
plot of  the figure aside.\\ \strut \\ \strut

 \end{minipage} 
 \end{tabular}
 \end{center}
 We invert eq.~(\ref{betainvert}) to extract $\Lambda_T$ as a function of
$\mu^2$  and then $\Lambda_{\overline{\rm MS}}$ from eq.~(\ref{ratTMS}).
Since $\Lambda_T$ and $\Lambda_{\overline{\rm
MS}}$ are constants independent  on the scale, the result should be a nice
``plateau'' if we were in the full perturbative regime. This is obviously not
the case as shown in black for  the three lattice spacings in the upper-right
and two down plots. We then use eq.~(\ref{ap-alphahNP}), fitting  $g^2(\mu^2)
\langle A^2 \rangle_{R,\mu_0^2}$ to have $\Lambda_{\overline{\rm MS}}$ as a nice
plateau. The result is shown in blue.  It looks quite convincing. {\bf From the
comparison of the black and blue curves it is evident that, although the fit has
been performed in the rather high-energy range 2.6-6.4, a sizeable non
perturbative correction is needed}. 
%\vskip -4cm

\begin{center}
\begin{tabular}{cc}
 \begin{minipage}{0.5\linewidth} \vspace {-4\baselineskip}
\includegraphics[width=6.5cm]{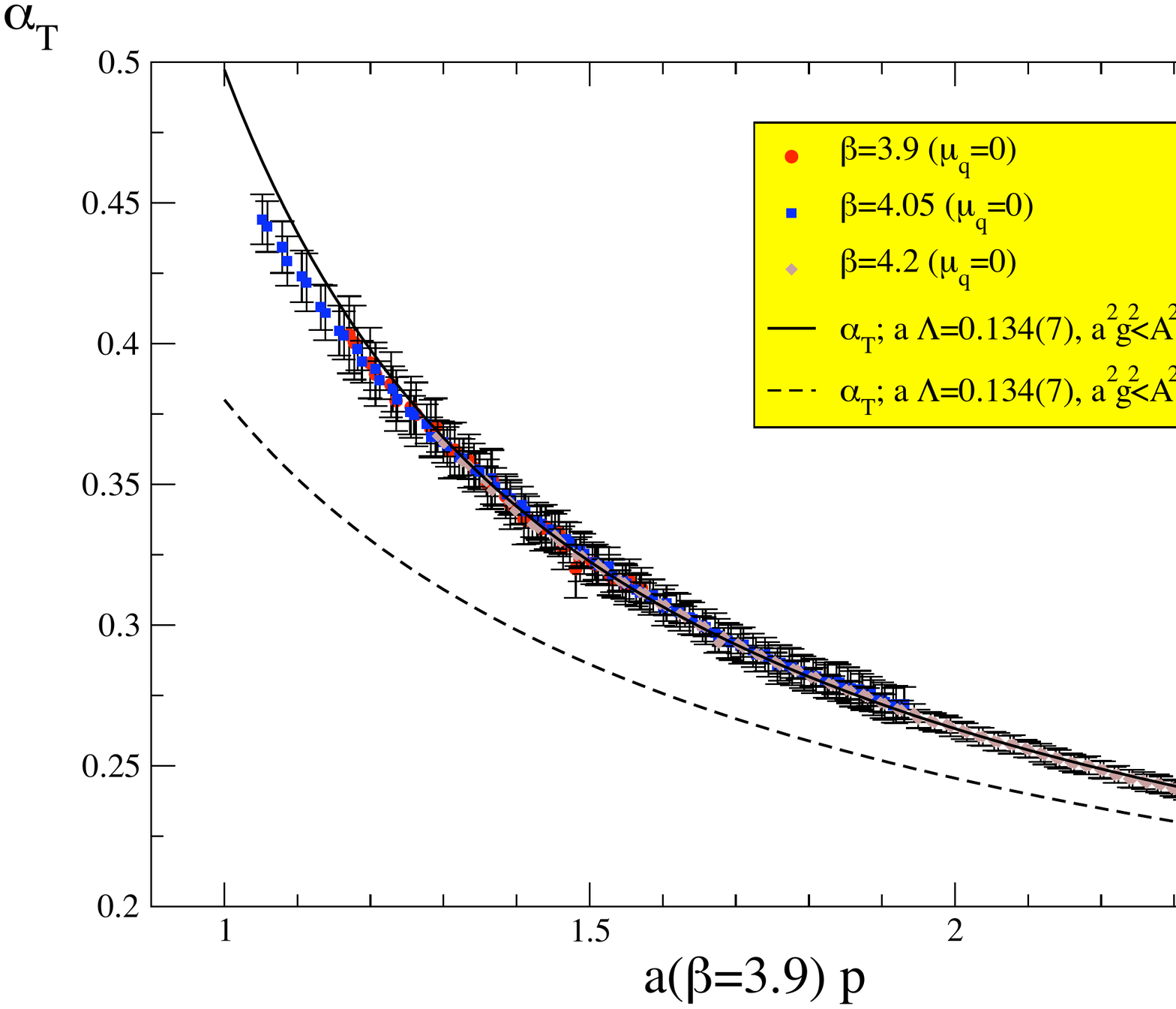}
 \end{minipage}
 \begin{minipage}{0.5\linewidth}
In the plot on the left we have merged the three curves of $\alpha_T(\mu^2)$ 
by fitting the lattice spacing ratios to have the best matching.
The compatibility of the three curves is quite impressive and the needed 
lattice spacing ratios agree very well with what is extracted from the
interquark potential~\cite{Boucaud:2008gn}. The fit has been performed both
using the leading logarithm~\cite{Boucaud:2008gn}
 (LL) and the three loop~\cite{Chetyrkin:2009kh} (${\cal O}(\alpha^4)$)
 formula for the Wilson coefficient. The result is:
 \end{minipage} 
 \end{tabular}
 \end{center}
 
 \bea\label{eq:finalalpha}
 &&\Lambda_{\overline{\rm MS}} = 330 \pm 32^{+0}_ {-33} \;{\rm MeV} \\
 &&g^2(\mu^2)\langle A^2 \rangle_{R,\mu} = 4.4 \pm 1.5 \pm 0.7\; {\rm GeV^2} ({\rm
 LL})\qquad 
 g^2(\mu^2)\langle A^2 \rangle_{R,\mu} = 2.7 \pm 1.0 \pm 0.7 {\rm GeV^2} ({\cal
 O} (\alpha^4))\nonumber 
 \eea
 
\section{The quark field renormalisation constant $Z_q$}
\begin{itemize}
\item
We compute the Fourier transform of the lattice quark propagator for
momentum $p$: it is a $12\times 12$ matrix $S(p)$. We define the renormalised
quark field
\bea
q_{\rm R} = Z_q\; q_{\rm bare}\qquad {\rm whence}\qquad Z_q(\mu^2=p^2) \equiv
\frac{-i}{12 p^2} {\rm Tr}\,\left[\frac{S_{\rm bare}^{-1} (p)\,\pslash}{p^2} \right]
\eea
\item We suppress the hypercubic artefacts which in this case are particularly 
large.
\item We perform the fit
\bea\label{eq:zqfit}
Z_{\rm qlatt} (p^2) = Z_{\rm pert} (p^2,\mu^2) \left (1 + \frac{C^Z_{\rm
wilson}(p^2,\mu^2)}{Z_{\rm pert} (p^2,\mu^2)} \langle A^2(\mu^2)\rangle \right)
+ c_{\rm a2p2}\; a^2 p^2
\eea
\end{itemize}
\begin{center}
\begin{tabular}{cc}
 \begin{minipage}{0.65\linewidth} 
\includegraphics[width=10cm]{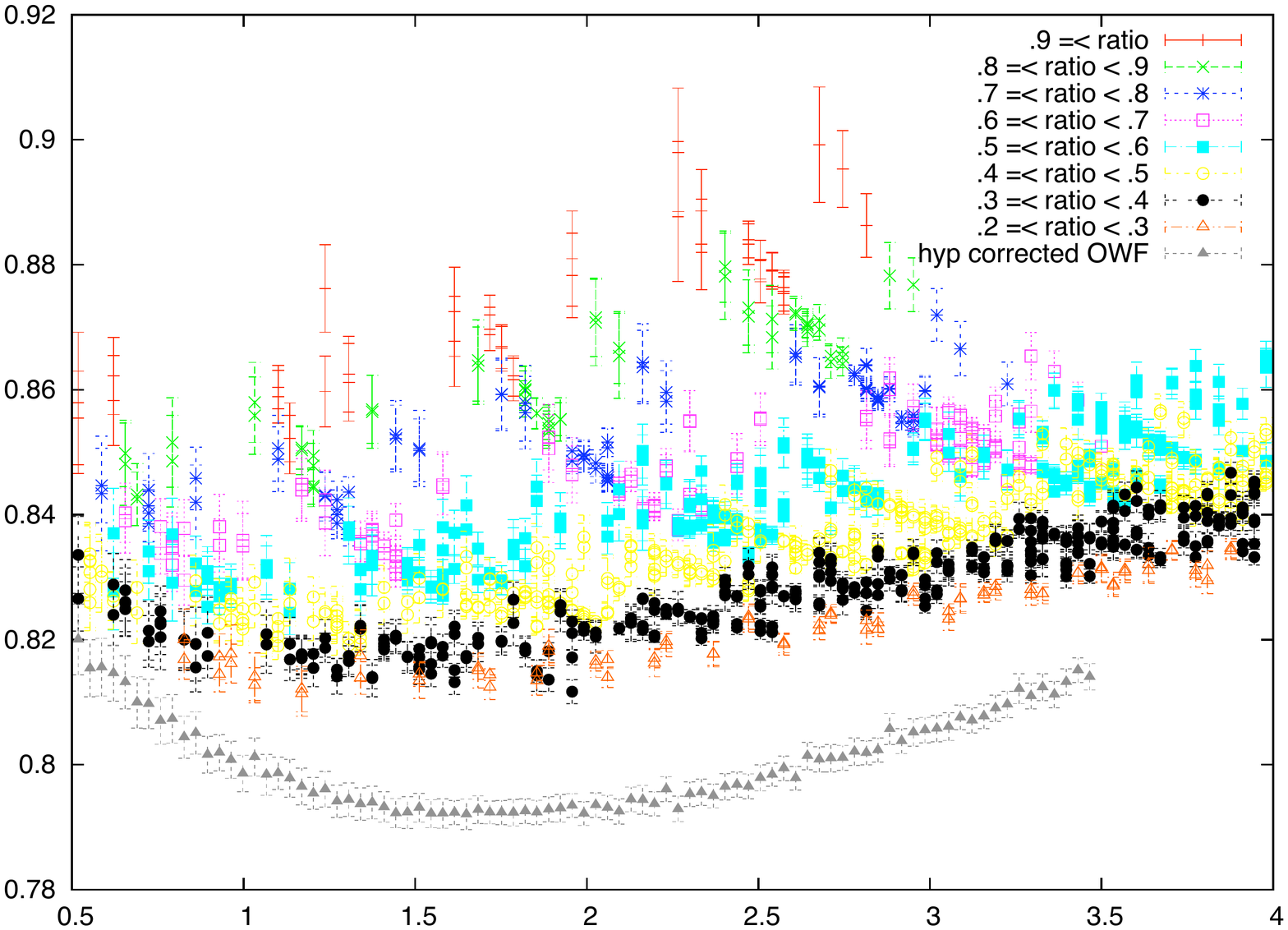}
 \end{minipage}
 \begin{minipage}{0.35\linewidth}
The plot on the left shows the result for $Z_{\rm qlatt}$ as a function of 
$p^2$ stemming directly from the lattice calculation. It is very far from 
the smooth dependence expected in the continuum. One sees a ``half-fishbone" 
structure which is a dramatic expression of hypercubic artefacts. Let us define
\bea
p^{[4]}=\sum_{\mu=1}^{4} p_\mu^4\qquad  {\rm ratio} \equiv \frac
{p^{[4]}}{(p^2)^2} \nonumber
\eea
 \end{minipage} 
 \end{tabular}
 \end{center}
 In the above plot the color code corresponds to the value of the parameter
 ``ratio" which is bounded $0.25 \le {\rm ratio} \le 1$. It is visible that the
 hypercubic artefacts increase with this  parameter. We reduce drastically these
 artefacts by an extrapolation down to  ratio$=0$~\cite{Boucaud:2005rm} :
  \bea\label{eq:owfexpan}
Z_q^{\mathrm {latt}}(a^2\,{p}^2, a^4p^{[4]}, a^6 p^{[6]}, ap_4, a^2\Lambda_{\rm
QCD}^2) &=& Z_q^{\mathrm {hyp\_corrected}}(a^2p^2,a^2\Lambda_{\rm QCD}^2) +
c_{a2p4} \; a^2 \frac{p^{[4]}}{p^2} \nonumber \\
&+& c_{a4p4} a^4\;p^{[4]}
\eea

\begin{figure}[hbt]
  \begin{center}\begin{tabular}{cc}{\hskip -1cm}
    \includegraphics[width=80mm]{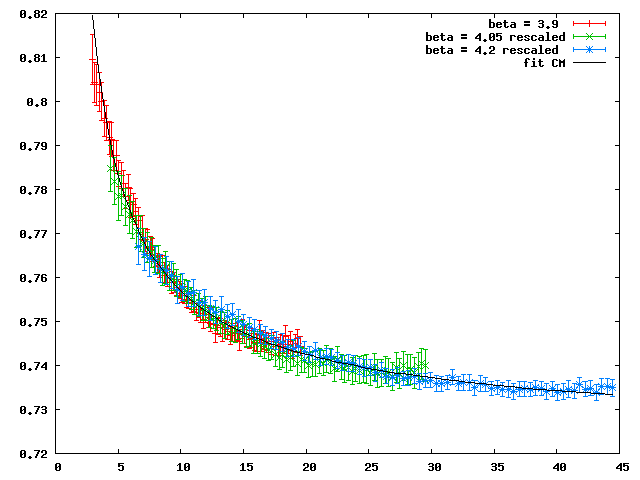}&
      \includegraphics[width=80mm]{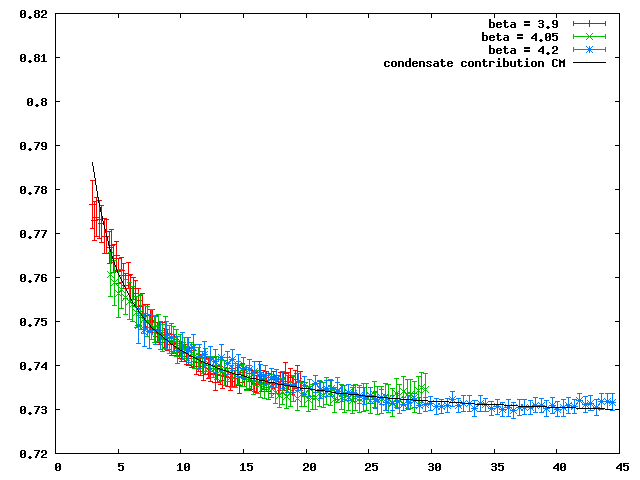}
%/homes/pene/lattice/ETMC/renormalisation/data_for_lat010/Zq/one_window     
\end{tabular} \end{center}
%\caption
{\small The merged plot  results with $\beta=4.05$ and
$\beta=4.2$ rescaled to the $\beta=3.9$.  The l.h.s shows  the
data corrected for all lattice artefacts. The r.h.s shows the same data
furthermore corrected by the perturbative running factor up to 10 GeV. The
horizontal axix is $p^2$ in GeV $^2$. The black line on the l.h.s corresponds to
the global fit with  perturbative running and the three-loops 
(Chetyrkin-Maier~\protect\cite{Chetyrkin:2009kh}) Wilson
coefficient for the  $1/p^2$ term. 
 The black line on the r.h.s corresponds only
to the $1/p^2$ times the three loops wilson coefficient added to  $Z_q^{\mathrm
{pert}}((10\, \mathrm {GeV})^2, 6/3.9)=0.726$}
\label{fig:all_betas}
\end{figure}

Once this non-perturbative hypercubic correction is performed, we merge
the results for three lattice spacings. The merged result is shown in  
the figure above l.h.s. In the r.h.s these data are corrected by the
 perturbative running factor up to 10 GeV. If non-perturbative
corrections were absent the curve should be flat. This is obviously not the
case. The results are shown in table~\ref{tab:p2andA2}.

%------------------------------------------------------------------------------
\begin{table}[h]
\centering
\begin{tabular}{||c|c|c|c|c|c||}
\hline
\hline
$\beta$ & $a^2$ fm$^2$ & $Z_q^{\mathrm{pert}}$ & $c_{a2p2}$ & 
$g^2 \langle{A^2}\rangle_{\mathrm tree }$ & $g^2 
\langle{A^2}\rangle_{\mathrm
CM}$ 
\\ \hline
%$3.9$ & 0.00689 & 0.742(5) &0.0178(13)&  5.40(36)  & 5.75(39)   \\ \hline
$3.9$ & 0.00689 & 0.726(5) &0.0201(13)&  3.20(38)  & 2.62(31)   \\ \hline
%$4.05$ &0.00456& 0.755(6) &0.0176(15)&  5.62(62)  &5.92(65)\\ \hline
$4.05$ &0.00456& 0.742(5) &0.0200(15)&  3.09(65)  &2.57(54)\\ \hline
%$4.2$ &0.00303&  0.770(3)  & 0.0172(8) &6.28(53) & 6.57(56) \\ \hline
$4.2$ &0.00303&  0.760(3)  & 0.0194(8) &3.23(55)  & 2.74(47) \\ \hline
average & & &  0.0201(3) &3.18(28) &  2.64(23) \\ \hline
\end{tabular}\label{tab:p2andA2}
\caption{Results for $Z_q^{\mathrm{pert}}$ (10GeV)  and 
$c_{a2p2}$~\protect\eq{eq:zqfit}  and the estimated $g^2 \langle A^2\rangle$ v.e.v from
the tree level $1/p^2$ term and  from the
Chetyrkin-Maier~\protect\cite{Chetyrkin:2009kh} (CM) Wilson coefficient.}
\end{table}

\section{conclusion}
 It is striking that 
the fitted value of $g^2 \langle{A^2}\rangle$ are independent of the lattice
spacings. It is even more striking that the values extracted from the quark 
renormalisation constant and the coupling constant are perfectly compatible.

This is very encouraging in the sense that all our criteria to identify really
a $g^2 \langle{A^2}\rangle$ condensate are fulfilled.

 \begin{table}[h]
\centering
\begin{tabular}{||c|c|c|c|c||}
\hline
\hline
  $N_f$ &  order 
$g^2 \langle{A^2}\rangle_{\mathrm tree }$ & $Z_q$ &$\alpha_T$ &3 gluons 
\\ \hline
 0  & LL &9.4(3) &  $5.2(1.1)$ & $10(3)$ \\ \hline 
 0 & $O(\alpha^4)$  & 9.0(3) & 3.7(8) &  \\ \hline 
 2 & LL &   2.7(4) & 4.4(1.6)& \\ \hline
 2& $O(\alpha^4)$ & 2.55(36)& $2.7(1.0)$ &\\ \hline
\end{tabular}
\caption{Comparison of estimates of $g^2 \langle{A^2}\rangle$ from different
quantities  at $N_f=0$ and $N_f=2$. All are taken at the scale $\mu= 10$\, GeV. 
LL means leading logarithm for the Wilson coefficient. $O(\alpha^4)$ refers to
Chetyrkin-Maier's~\protect\cite{Chetyrkin:2009kh} computation. }
\label{tab:global_comp}
\end{table}

In table~\ref{tab:global_comp} we perform a global comparison of the estimates
of  $g^2(\mu^2) \langle{A^2}\rangle_{\mu}$ both in the quenched and unquenched
case. The agreement between $g^2(\mu^2) \langle{A^2}\rangle_{\mu}$ estimated
from different observables is not very good in the quenched case. This does not 
induce in our mind any doubt about the existence of a   $g^2(\mu^2)
\langle{A^2}\rangle_{\mu}$ ($\mu= 10$ GeV) condensate in Landau gauge for the
following reason: we have performed a large number of fits with different
inputs, and we have never found $g^2(\mu^2) \langle{A^2}\rangle_{\mu}$
compatible with zero at better than four sigmas. On the other hand the fitting
procedure is certainly delicate due to many correlations between  $g^2(\mu^2)
\langle{A^2}\rangle_{\mu}$ and lattice artefacts. Our final conclusion is {\bf
that the accurate estimate of its value needs some  improvements in our fitting
method due to several correlations difficult to disentangle. There are however
strong evidences in favor of the existence of a positive condensate 
 $g^2(\mu^2) \langle{A^2}\rangle_{\mu}$
in the range 2-10 GeV$^2$ in the $\overline MS$ scheme at
$\mu=10$ GeV}.

%%%%%%%%%%%%%%%%%%%%%%%%%%%%%%%%%%%%%%%%%%%%%%%%%%%

\end{document}